\documentclass[aps,prc,twocolumn,floatfix,nofootinbib,showpacs,superscriptaddress]{revtex4-1}

\usepackage[mathscr,scaled=1.15]{urwchancal}
\DeclareFontFamily{OT1}{pzc}{}
\DeclareFontShape{OT1}{pzc}{m}{it}%
{<-> s * [1.15] pzcmi7t}{}
\DeclareMathAlphabet{\mathpzc}{OT1}{pzc}{m}{it}

\usepackage{amsmath,amsfonts,amssymb,bm}
\usepackage{graphicx}
\usepackage{color}

\definecolor{purple}{rgb}{0.5,0,0.5}
\definecolor{blue}{rgb}{0.0,0,0.9}

\begin{document}

\title{Zero mode in a strongly coupled quark gluon plasma}
\email[Communicating authors: ]{yxliu@pku.edu.cn, cdroberts@anl.gov} 

\author{Fei Gao}
\affiliation{Department of Physics and State
Key Laboratory of Nuclear Physics and Technology, Peking
University, Beijing 100871, China}
\affiliation{Collaborative Innovation Center of Quantum Matter, Beijing 100871, China}

\author{Si-Xue Qin }
\affiliation{Institut f\"{u}r Theoretische Physik, Johann
Wolfgang Goethe University, Max-von-Laue-Str.\,1, D-60438 Frankfurt am Main, Germany}

\author{Yu-Xin Liu }
\affiliation{Department of Physics and State Key Laboratory of
Nuclear Physics and Technology, Peking University, Beijing 100871,
China}
\affiliation{Collaborative Innovation Center of Quantum Matter, Beijing 100871, China}
\affiliation{Center for High Energy Physics, Peking University, Beijing 100871, China}

\author{Craig D.~Roberts}
\affiliation{Division of Physics, Argonne National Laboratory, Argonne, IL 60439, USA}

\author{Sebastian M.~Schmidt}
\affiliation{Institute for Advanced Simulation, Forschungszentrum J\"ulich and JARA, D-52425 J\"ulich, Germany}

\date{9 January 2014}

\begin{abstract}
In connection with massless two-flavour QCD, we analyse the chiral symmetry restoring phase transition using three distinct gluon-quark vertices and two different assumptions about the long-range part of the quark-quark interaction.  In each case, we solve the gap equation, locate the transition temperature $T_c$, and use the maximum entropy method to extract the dressed-quark spectral function at $T>T_c$.
Our best estimate for the chiral transition temperature is $T_c=147\pm 8\,$MeV; and the deconfinement transition is coincident.
For temperatures markedly above $T_c$, we find a spectral density that is consistent with those produced using a hard thermal loop expansion, exhibiting both a normal and plasmino mode.
On a domain $T\in [T_c,T_s]$, with $T_s \simeq 1.5 T_c$, however, with each of the six kernels we considered, the spectral function contains a significant additional feature.  Namely, it displays a third peak, associated with a zero mode, which is essentially nonperturbative in origin and dominates the spectral function at $T=T_c$.
We suggest that the existence of this mode is a signal for the formation of a strongly-coupled quark-gluon plasma and that this strongly-interacting state of matter is likely a distinctive feature of the QCD phase transition.
\end{abstract}


\maketitle

\section{Introduction}
It is widely held that experiments at the relativistic heavy ion collider (RHIC) have produced a quark-gluon plasma (QGP) \cite{Adcox:2004mh,Arsene:2004fa,Back:2004je,Adams:2005dq}.  Analyses of RHIC experiments \cite{Song:2008hj,Song:2010mg}, which couple viscous fluid dynamics for the QGP with a microscopic transport model for hadronic freeze-out, have related the measured elliptic flow, $v_2$, to the ratio $\eta/s$, where $\eta$, $s$ are, respectively, the medium's shear viscosity and entropy density.  Such studies yield $1 < 4\pi \eta/s < 2.5$ on the domain $1\lesssim T/T_c \lesssim 2$, where $T_c$ is the temperature required for QGP creation.

To place this result in context we note that an ideal fluid has $\eta=0$, and hence no resistance to the appearance and growth of transverse velocity gradients.  A perfect fluid with near-zero viscosity is the best achievable approximation to that ideal.  Arguments within string theory have been used to suggest that in gauge theories with a gravity dual one has a lower bound on viscosity; viz. \cite{Kovtun:2004de}, $1\leq 4\pi\eta/s$.  The RHIC result above has therefore led many to conclude that the QGP is an almost perfect fluid on $1\lesssim T/T_c \lesssim 2$; i.e., it as close as physically achievable to the case of zero viscosity.  Considering Newton's law for viscous fluid flow; viz., $(v/d) = (1/\eta) (F/A)$, it is apparent that in near-perfect fluids a macroscopic velocity gradient is achieved from a microscopically small pressure.  Strong interactions between particles constituting the fluid are necessary to achieve this outcome.  Hence the medium produced at RHIC is commonly described as a strongly-coupled quark gluon plasma (sQGP), in which case its properties should differ substantially from those anticipated via perturbation theory.

Quantum chromodynamics (QCD) is known to produce the bulk of the mass of normal matter \cite{national2012Nuclear,Cloet:2013jya}.   The $T=0$ theory is characterised by confinement and dynamical chiral symmetry breaking (DCSB), phenomena that are represented by a range of order parameters, some or all of which vanish in the sQGP.  Understanding the sQGP therefore requires elucidation of the behaviour and properties of quarks and gluons within this state.  Perturbative techniques have been developed for use far above $T_c$; viz., the hard thermal loop (HTL) expansion \cite{Pisarski:1988vd,Braaten:1990wp}, which has enabled the computation of gluon and quark thermal masses $m_T\sim g T$ and damping rates $\gamma_T\sim g^2 T$, with $g=g(T)$ being the strong running coupling.  It also suggests the existence of a collective plasmino or ``abnormal'' branch to the dressed-quark dispersion relation, which is characterised by antiparticle-like evolution at small momenta \cite{Blaizot:2001nr}.

Owing to asymptotic freedom, the running coupling in QCD increases as $T\to T_c^+$.  Therefore, a simple interpretation of the HTL results suggests the plasmino should disappear before $T_c$ is reached from above because $\gamma_T$ increases more rapidly than $m_T$ and $\gamma_T/m_T\sim 1$ invalidates a quasiparticle picture.
On the other hand, nonperturbative studies using lattice-regularised quenched-QCD  \cite{Karsch:2009tp} or Dyson-Schwinger equations (DSEs) \cite{Qin:2010pc} suggest that the plasmino branch persists in the vicinity of $T_c$.

Resolving the active degrees of freedom in the neighbourhood of $T_c$ is important because the spectral properties of the dressed-quark propagator are intimately linked with light-quark confinement \cite{Gribov:1999ui,Munczek:1983dx,Stingl:1983pt,Cahill:1988zi,Krein:1990sf,%
Efimov:1993zg,Dokshitzer:2004ie,Roberts:2007ji,Bashir:2008fk,Cloet:2013jya} and it is long-range modes which might produce strong correlations.  Further in this connection, there is an accumulating body of evidence that, in addition to the normal and plasmino modes, an essentially nonperturbative fermionic zero mode exists on a material domain of $T>T_c$ \cite{Kitazawa:2005mp,Kitazawa:2007ep,Harada:2008vk,Qin:2010pc,Qin:2013ufa} and, moreover, that the domain upon which it exists defines the extent of the sQGP.

Hitherto, however, those studies which expose the zero mode have worked in the simplest symmetry-preserving approximation of the DSEs; i.e., they have effectively used the rainbow-ladder (RL) truncation.  That truncation may be characterised as incorporating dressing only for the $\gamma_\mu$ component of the gluon-quark vertex  \cite{Munczek:1994zz,Bender:1996bb,Maris:1997tm,Maris:1999nt}.  It is accurate for ground-state vector- and isospin-nonzero-pseudoscalar-mesons \cite{Maris:2003vk,Chang:2011vu,Bashir:2012fs}, and nucleon and $\Delta$ properties \cite{Eichmann:2011ej,Chen:2012qr,Segovia:2013rca,Segovia:2013uga} because corrections in these channels largely cancel, owing to parameter-free preservation of the Ward-Green-Takahashi identities \cite{Ward:1950xp,Green:1953te,Takahashi:1957xn}.  However, they do not cancel in other channels \cite{Roberts:1996jx,Roberts:1997vs,Bender:2002as,Bhagwat:2004hn}.  Hence studies based on the RL truncation, or low-order improvements thereof, have usually provided poor results for scalar- and axial-vector-mesons \cite{Cloet:2007pi,Burden:1996nh,Watson:2004kd,Maris:2006ea,Fischer:2009jm,%
Krassnigg:2009zh}, produced masses for exotic states that are too low in comparison with other estimates \cite{Cloet:2007pi,Qin:2011dd,Burden:1996nh,Qin:2011dd,Qin:2011xq,Krassnigg:2009zh}, and exhibit gross sensitivity to model parameters for excited states \cite{Qin:2011dd,Qin:2011xq,Holl:2004fr,Holl:2004un} and tensor-mesons \cite{Krassnigg:2010mh}.  In these circumstances one must conclude that physics important to these states is omitted.

Given these observations, it is evidently worth analysing both the impact of more sophisticated gluon-quark vertices on the $T\neq 0$ phase transitions, and whether they affect the existence and appearance of the normal, plasmino and zero modes.  Likewise, one should determine the impact of different assumptions about the long-range part of the quark-quark interaction.  To that end, herein we employ three different vertex \emph{Ans\"atze} -- the rainbow, Ball-Chiu (BC) \cite{Ball:1980ay} and anomalous magnetic moment-improved (DB) \cite{Kochelev:1996pv,Bicudo:1998qb,Diakonov:2002fq,Chang:2010hb,Chang:2011ei,Bashir:2011dp,%
Qin:2013mta} structures; and two different interactions -- Maris-Tandy (MT) \cite{Maris:1997tm,Maris:1999nt} and Qin-Chang (QC) \cite{Qin:2011dd,Qin:2011xq}.  With each of the six permutations that these inputs allow, we solve the gap equation, locate the transition temperature $T_c$, and use the maximum entropy method (MEM) to extract the dressed-quark spectral function at $T>T_c$.  The latter contains the information needed in order to expose the properties of fermionic modes in the sQGP.

Our report is arranged as follows.  Section\,\ref{secTwo} describes the quark gap equation, the vertex \emph{Ans\"atze} and the interaction models.  Section\,\ref{secMEM} recapitulates upon the MEM and explains our use of the method.  Numerical results are explained in Sec.\,\ref{secResults}; and we summarise in Sec.\,\ref{secEnd}.

\section{Quark DSE}
\label{secTwo}
At nonzero temperature, the gap equation is \cite{Roberts:2000aa}
\begin{eqnarray}
\nonumber
\lefteqn{S(\vec{p},i\omega_n)^{-1} = Z_{2}^{A} i \vec{\gamma} \cdot \vec{p} + Z_{2}^{} i \gamma_{4}^{} \omega_{n}^{}} \\
& +&    \, T\! \sum_{l} \frac{4}{3} \int \frac{d^3 q}{(2\pi)^3}
g^2 D_{\mu\nu}(\vec{p} \! - \! \vec{q},{\omega_{n}}\! - \! {\omega_{l}})\notag    \\
& & \times(Z_1^A\gamma_{\mu}^T+Z_1 \gamma_{\mu}^L)
S(\vec{q},i\omega_l)\Gamma_\nu(\vec{q},\omega_{l},\vec{p},
\omega_n),\label{eq:gapeq}
\end{eqnarray}
where $S(\vec{p},i\omega_n)$ is the dressed-quark propagator, with $\omega_n=(2n + 1){\pi} T$, $n\in \mathbb{Z}$ being the fermion Matsubara frequency; $D_{\mu\nu}^{}$ is the dressed-gluon propagator; $\Gamma_{\nu}$ is the dressed-quark-gluon vertex; $Z_{1,2}^{}$, $Z_{1,2}^{A}$ are, respectively, vertex and wave function renormalisation constants; and $\gamma^{T}_{\mu} = \gamma_{\mu}^{} - \gamma_{\mu}^{L}$, with $\gamma^{L}_{\mu} = u_{\mu}^{} \gamma_{\alpha}^{} u_{\alpha}^{}$ and $u=(0,0,0,1)$.  We use the same renormalisation scheme and scale as Refs.\,\cite{Bender:1996bm,Maris:2000ig}.

The solution of Eq.\,\eqref{eq:gapeq} has the form
\begin{eqnarray}
\nonumber
\lefteqn{S(\vec{p},i\omega_n)^{-1}}\\
& =&  i\vec{\gamma} \cdot \vec{p}\, A(\vec{p}\,^2,\omega_n) + i\gamma_4\omega_n C( \vec{p}\,^2,\omega_n,) + B(\vec{p}\,^2,\omega_n,)\, . \quad
\label{eq:qdirac}
\end{eqnarray}
The chiral limit is defined by $\hat m=0$, where $\hat m$ is the renormalisation group invariant current-quark mass.  If $B(\vec{p}\,^2,\omega_n)\neq 0$ in that limit, then chiral symmetry is dynamically broken and the symmetry is realised in the Nambu mode.  The critical temperature for chiral symmetry restoration is that value of $T=T_c$ above which $B(\vec{p}\,^2,\omega_n) \equiv 0$ is the only solution to Eq.\,\eqref{eq:qdirac}.  This situation defines the Wigner phase.  In DSE studies conducted to date, the critical temperature for chiral symmetry restoration typically coincides with that for deconfinement \cite{Bender:1996bm,Blaschke:1997bj,Maris:2000ig,Roberts:2000aa,Fischer:2009gk,%
Qin:2010pc,Luecker:2012hf,Wang:2013wk,Qin:2013ufa}.

\subsection{Vertices}
\label{vertices}
The gap equation is defined once the elements in its kernel are specified.  A simple kernel is obtained using the rainbow truncation
\begin{equation}
\label{Vrainbow}
\Gamma_\mu^R(\vec{q},\omega_{l},\vec{p},\omega_{n}) = \gamma_\mu\,,
\end{equation}
with all associated dressing of this Dirac matrix structure absorbed into the interaction $g^2 D_{\mu\nu}({\omega_{n}}\! - \! {\omega_{l}})$ \cite{Maris:1997tm,Bhagwat:2003vw}.

A more sophisticated \emph{Ansatz}, which includes some aspects of DCSB in the vertex, is obtained with \cite{Ball:1980ay}
\begin{eqnarray}
\nonumber\lefteqn{\Gamma_{\mu}^{BC}(\vec{q},\omega_{l},\vec{p},\omega_{n}) = \gamma^T_{\mu}\Sigma_A+\gamma^L_{\mu}\Sigma_C + (p_n+q_l)_\mu} \\
 &&\times [\frac{1}{2}\gamma^T_\alpha(p_n+q_l)_\alpha\Delta_A +\frac{1}{2}\gamma^L_\alpha(p_n+q_l)_\alpha\Delta_C-i\Delta_B] \, ,
\label{eq:BCvertex}
\end{eqnarray}
with ($F= A, B, C$)
\begin{eqnarray}
& & p_n=(\vec{p},\omega_n), \quad q_l =(\vec{q},\omega_l) ,  \notag \\
& & \Sigma_F(\vec{q}^{\,2},\omega^2_l,\vec{p}^{\,2},\omega^2_n) =\frac{1}{2}[F(\vec{q}^{\,2},\omega^2_l)+F(\vec{p}^{\,2},\omega^2_n)]\,, \notag\\
& & \Delta_F(\vec{q}^{\,2},\omega^2_l,\vec{p}^{\,2},\omega^2_n)  =  \frac{F(\vec{q}^{\,2},\omega^2_l)-F(\vec{p}^{\,2},\omega^2_n)}{q^2_l-p^2_n} \, .
 \end{eqnarray}

It is shown elsewhere \cite{Kochelev:1996pv,Bicudo:1998qb,Diakonov:2002fq,Chang:2010hb,Chang:2011ei,%
Bashir:2011dp,Qin:2013mta} that Eq.\,\eqref{eq:BCvertex} is incomplete in the presence of DCSB.  A dressed-quark anomalous chromomagnetic moment term must also be included.  Defining $\sigma_{\mu \nu}^{} = (i/2)[\gamma_{\mu}^{} , \gamma_{\nu}^{} ]$,  $T_{\mu\nu}=\delta_{\mu\nu}-k_{\mu}k_{\nu}/k^{2}$, one can express this improvement as
\begin{eqnarray}
 \label{eq:ACMvertex}
\Gamma_{\mu}^{DB} &=& \Gamma_{\mu}^{BC}+\Gamma_{\mu}^{ACM} \, , \\
\Gamma_{\mu}^{ACM}&=&\Gamma_{\mu}^{ACM_{4}}+\Gamma_{\mu}^{ACM_{5}} \, , \\
\Gamma_{\mu}^{ACM_4}&=&[T_{\mu\nu} l_{\nu} \gamma \cdot k + i T_{\mu\nu}\gamma_{\nu}\sigma_{\rho\sigma}l_{\rho}k_{\sigma}]\tau_4(p,q) \, , \\
\Gamma_{\mu}^{ACM_5}&=&\sigma_{\mu\nu}k_{\nu}\tau_5(p,q) \, , \\
\tau_4&=&\frac{4\tau_5(p,q)[M(p^{2})+M(q^{2})]}{p^{2}+M(p^{2})^2+q^{2}+M(q^{2})^2} \, ,
 \end{eqnarray}
where $k=p-q$, $l = (p+q)/2$, $\tau_{5}^{} =\varsigma\Delta_{B} $ with $\varsigma=0.65$, and $M(x)=B(x)/A(x)$.  N.B.\,For $T\geq T_c$, $\Gamma_\mu^{DB} = \Gamma_\mu^{BC}$; but this does not entail identical results for BC and DB vertices at $T>T_c$ because a realistic description of in-vacuum physics requires a different value of the interaction strength in each case, as discussed below in connection with Eq.\,\eqref{eqInteractionParameters}.

\subsection{Interactions}
\label{secInteractions}
A widely used form for the interaction $g^2D_{\mu\nu}$ is that presented in Refs.\,\cite{Maris:1999nt}.  However, as explained elsewhere \cite{Qin:2011dd,Qin:2011xq}, some aspects of the infrared behaviour of that interaction, notably its zero at $k^2=0$, are in stark conflict with the results of modern DSE and lattice studies.  Those analyses indicate that the gluon propagator is a bounded, regular function of spacelike momenta, which achieves its maximum value on this domain at $k^2=0$
\cite{Bowman:2004jm,Bogolubsky:2009dc,Boucaud:2010gr,Oliveira:2010xc,Cucchieri:2011ig,Aguilar:2012rz,Ayala:2012pb,Dudal:2012zx,%
Strauss:2012dg,Weber:2012vf,Zwanziger:2012xg,Blossier:2013te}; and the dressed-quark-gluon vertex does not possess any structure which can qualitatively alter this behaviour \cite{Skullerud:2003qu,Bhagwat:2004kj}.  The interaction in Refs.\,\cite{Qin:2011dd,Qin:2011xq} expresses these features.

Herein we consider both types of interaction, which at nonzero temperature can be represented in the form \cite{Maris:2000ig}
\begin{eqnarray}
\nonumber
& & g^{2} D_{\mu\nu}(\vec{p}-\vec{q},\omega_{n} - \omega_{l})
\Gamma_{\nu}(\vec{q},{\omega_{l}},\vec{p},{\omega_{n}}) \\
& = & [{P_{T}^{\mu\nu}}({k_{\Omega}}){D_{T}}({k_{\Omega}})
+{P_{L}^{\mu\nu}}({k_{\Omega}}){D_{L}}({k_{\Omega}})]{\Gamma_{\nu}}, \label{eq:model}
\end{eqnarray}
where $\Gamma_{\nu}$ is one of the \emph{Ans\"atze} in Sec.\,\ref{vertices}; ${k_{\Omega}}:=(\vec{k},\Omega)=(\vec{p}-\vec{q},{\omega_{n}} -{\omega_{l}})$;
\begin{eqnarray}
P_T^{\mu\nu}(k_{\Omega})=\left\{
\begin{aligned}
&0, \qquad \qquad \quad {\mu}\text{ and/or } {\nu} = 4 \, ,  \\
&\delta_{ij}-\frac{{k_{i}} {k_{j}}}{k^2}, \quad {\mu}, {\nu} =1,2,3
\, ,
\end{aligned}
\right.
\end{eqnarray}
with $P_{L}^{\mu\nu} + P_{T}^{\mu\nu} = \delta_{\mu\nu} - k_\Omega^\mu k_\Omega^\nu /k_\Omega^2$; and
\begin{eqnarray}
{D_{T}(k_{\Omega})} &=&\mathcal{D}({k^{2}_{\Omega}},0), \quad
{D_{L}(k_{\Omega})} =\mathcal{D}({k^{2}_{\Omega}},{m_{g}^{2}})\,,
\end{eqnarray}
where $s_{\Omega}^{} =k^{2}_{\Omega} = \Omega^2 + \vec{k}\,^2 + m^{2}_{g}$ and, in generalising to $T\neq 0$, we have followed perturbation theory and included a Debye mass in the longitudinal part of the gluon propagator: $m_g^2= (16/5) T^2$.

Following this procedure, the interactions in Refs.\,\cite{Maris:1999nt,Qin:2011dd} are represented via
\begin{eqnarray}
\nonumber
\mathcal{D}({k^{2}_{\Omega}}, {m_{g}^{2}}) & = & \mathcal{D}_{\rm ir}({k^{2}_{\Omega},m_g^2})\\
& & + \frac{8{\pi^{2}} {\gamma_{m}}}{{\ln}[ \tau \! + \! (1 \! + \!
{s_{\Omega}^{}}/{\Lambda_{\text{QCD}}^{2}} ) ^{2} ] } \,
{\cal F}(s_{\Omega}^{}) \, ,
\end{eqnarray}
with ${\cal F}(s_{\Omega}) = (1-\exp(-s_{\Omega}/4 m_{t}^{2})/s_{\Omega}$, , $\tau=e^2-1$, $m_t=0.5\,$GeV, $\gamma_m=12/25$, $\Lambda^{N_f=4}_{\text{QCD}}=0.234\,$GeV; and, respectively,
\begin{eqnarray}
\label{DMT}
\mathcal{D}_{\rm ir}^{MT}({k^{2}_{\Omega},m_g^2}) &=& 4{\pi^{2}} D
\frac{s_{\Omega}^{}}{\mathpzc{w}^{6}} e^{-{s_{\Omega}^{}}/\mathpzc{w}^{2}}\,,\\
\label{DQC}
\mathcal{D}_{\rm ir}^{QC}({k^{2}_{\Omega},m_g^2}) &=& 8{\pi^{2}} D
\frac{1}{\mathpzc{w}^{4}} e^{-{s_{\Omega}^{}}/\mathpzc{w}^{2}}\,,
\end{eqnarray}
where, as we now explain, $\mathpzc{w}$ is a parameter and $\mathpzc{w} D=$ fixed.

\subsection{Qualitative features of the kernels}
\label{secQualitative}
It is worth reiterating that at $T=0$ each of the six kernels constructed from the elements expressed in Secs.\,\ref{vertices}, \ref{secInteractions} reproduce the results of perturbative QCD for $p^2\gtrsim 2\,$GeV$^2$, so any model-dependence is restricted to the infrared.  The single model parameter, $\mathpzc{w}$, is fixed by requiring that a particular kernel provides the best possible description of a diverse array of ground-state meson observables \cite{Maris:1999nt,Qin:2011dd,Qin:2011xq,Chang:2011ei}.  As made apparent in Refs.\,\cite{Maris:2002mt,Eichmann:2008ae,Qin:2011dd,Qin:2011xq}, there is typically a material domain of $\mathpzc{w}$ within which the predicted value for observables is unchanged so long as $\mathpzc{w} D = \,$constant.  Herein we primarily use the following values, in GeV:
\begin{equation}
\label{eqInteractionParameters}
\begin{array}{lccc}
    & {\rm rainbow} & BC & DB \\
MT  \quad \begin{array}{c}
    (\mathpzc{w} D)^{1/3} \\ \mathpzc{w} \end{array}
    &  \begin{array}{l} 0.72 \\ 0.4 \end{array}    
    & \begin{array}{l}  0.54 \\ 0.4 \end{array}    
    & \begin{array}{l}  0.40 \\ 0.4 \end{array} \\ 
QC  \quad \begin{array}{c}
    (\mathpzc{w} D)^{1/3} \\ \mathpzc{w} \end{array}
    & \begin{array}{l} 0.80 \\ 0.5 \end{array}     
    & \begin{array}{l} 0.60 \\ 0.5 \end{array}     
    & \begin{array}{l} 0.53 \\ 0.5 \end{array}     
\end{array}\,,
\end{equation}
which are those determined elsewhere \cite{Maris:1999nt,Qin:2011dd,Qin:2011xq,Chang:2011ei}, modified slightly, if necessary, so as to ensure a uniform result for the chiral-limit value of the in-pion condensate \cite{Maris:1997tm,Brodsky:2009zd,Brodsky:2010xf,Chang:2011mu,Brodsky:2012ku,Chang:2013epa}: $\langle \bar q q \rangle_{1\,{\rm GeV}}^\pi = (-0.24\,{\rm GeV})^3$.

It is notable that the extension of these interaction kernels to $T>0$ preserves the agreement with perturbative QCD at large spacelike momenta.  However, an insufficiency of the interactions is that $D$, the parameter expressing their infrared strength for fixed $\mathpzc{w}$, is assumed to be $T$-independent.  Since the nonperturbative part of the interaction should be screened for $T\gtrsim T_c$, we remedy that by writing $D\to D(T)$ with
\begin{eqnarray}
D(T)= D \left\{
\begin{array}{ll}
\displaystyle
1 \,, &   T<T_{\text{p}} \, ,  \\
\displaystyle
\frac{a}{b+ \ln[T/\Lambda_{QCD}]}\,, &  T \ge
T_{\text{p}}
\end{array}
\right.\,, \label{DTfunction}
\end{eqnarray}
where $T_{\rm p}$ is a ``persistence'' temperature; i.e., a scale below which nonperturbative effects associated with confinement and dynamical chiral symmetry breaking are not materially influenced by thermal screening.  Logarithmic screening is typical of QCD and we take $T_{\rm p}=T_c$ herein.
The parameters $a$, $b$ are fixed by applying the constraint $m_T = 0.8 \, T$ for $T \gtrsim 2\, T_c$; viz., requiring a thermal quark mass consistent with lattice-QCD \cite{Karsch:2009tp}, a procedure which yields:
\begin{equation}
\begin{array}{lccc}
    & {\rm rainbow} & BC & DB \\
MT  \quad \begin{array}{r}
    10 a \\ b \end{array}
    &  \begin{array}{l} 0.30 \\ 0.56 \end{array}
    & \begin{array}{l}  0.30 \\ 0.41 \end{array}
    & \begin{array}{l}  0.25 \\ 0.62 \end{array}\\
QC  \quad \begin{array}{r}
    10 a \\ b \end{array}
    & \begin{array}{l} 0.29 \\ 0.53 \end{array}
    & \begin{array}{l} 0.53 \\ 0.35 \end{array}
    & \begin{array}{l} 0.57 \\ 0.58 \end{array}
\end{array}\,.
\end{equation}

\section{Maximum Entropy Method}
\label{secMEM}
The solution of Eq.\,\eqref{eq:gapeq} can be used to compute the retarded real-time propagator
\begin{equation}
\label{eq:qreal}
S^R(\vec{p},\omega)= \left. S(\vec{p},i{\omega_{n}})\right|_{i{\omega_{n}}
\rightarrow\omega+i\eta^+}\,,
\end{equation}
from which one may obtain the spectral density
\begin{equation}
\label{eq:spec}
\rho(\vec{p},\omega)=-2 \Im\,S^R(\vec{p},\omega) \, .
\end{equation}
Equations~(\ref{eq:qreal}) and (\ref{eq:spec}) are equivalent to the statement:
\begin{equation}
S(\vec{p},i\omega_n) = \frac{1}{2\pi}\int_{-\infty}^{+\infty}\!\!\!\!\!\!
d\omega^\prime\,\frac{\rho(\vec{p},\omega^\prime)}{\omega^\prime-i{\omega_{n}}
} \, . \label{eq:mat_spec}
\end{equation}
N.B.\ If one requires a nonnegative spectral density, then Eq.\,(\ref{eq:mat_spec}) is only valid on the deconfined domain.

In the absence of DCSB, the spectral density associated with the propagator in Eq.\,(\ref{eq:qdirac}) can be expressed
\begin{eqnarray}
\label{eq:decomposition-rho}
\rho(\vec{p},\omega) = {\rho_{+}} (|\vec{p}|,\omega ) {P_{+}} +
{\rho_{-}} (|\vec{p}|,\omega) {P_{-}}  \,,
\end{eqnarray}
where $P_\pm=(\gamma_{4} \pm i\vec{\gamma}\cdot \vec{u}_p)/2$, $\vec{u}_p \cdot \vec{p} = |\vec{p}|$, are operators which project onto spinors with a positive or negative value for the ratio ${\cal H}:=\,$helicity/chirality: ${\cal H} = 1$ for a free positive-energy fermion.  By analogy with Eq.\,\eqref{eq:decomposition-rho}, one may write
\begin{equation}
\label{eq:decomposition-S}
S(\vec{p},\omega) = S_+(|\vec{p}|,\omega ) {P_{+}} +
S_-(|\vec{p}|,\omega) {P_{-}}  \,.
\end{equation}

The spectral density is interesting and expressive because it reveals the manner by which interactions distribute the single-particle spectral strength over momentum modes; and the behaviour at $T\neq 0$ shows how that is altered by a heat bath. As with many useful quantities, however, it is nontrivial to evaluate $\rho(|\vec{p}|,\omega)$.  Nonetheless, if one has at hand a precise numerical determination of the dressed-quark propagator in Eq.\,(\ref{eq:qdirac}), then it is possible to obtain the spectral density via the maximum entropy method (MEM) \cite{Nickel:2006mm}, as we shall shortly explain.

The dressed-quark propagator is a matrix-valued complex function.  Further analysis can therefore be simplified by employing a Fourier transform of Eq.\,(\ref{eq:mat_spec}).  Using the identity
\begin{equation}
T\sum_n\frac{e^{-i\omega_n\tau}}{\omega-i\omega_n}=\frac{e^{-\omega\tau}}{1+e^{-\omega/T}},
\end{equation}
combined with Eqs.\,(\ref{eq:decomposition-rho}) and (\ref{eq:decomposition-S}), then Eq.\,(\ref{eq:mat_spec}) entails
\begin{eqnarray}
\label{eqDrhoO}
D_\pm(|\vec{p}|,\tau)&:=&T\sum_{n}e^{-i\omega_n\tau}S_\pm(|\vec{p}|,i\omega_n)  \\
 &=&\int^{+\infty}_{-\infty}\frac{d\omega}{2\pi} \, \rho_\pm(|\vec{p}|,\omega)(\frac{e^{-\omega\tau}}{1+e^{-\omega/T}})\,.
\label{eq:Drho}
\end{eqnarray}
It is in connection with Eq.\,\eqref{eq:Drho}, which defines real-valued functions, that the MEM may be used effectively.

The MEM \cite{Bryan:1990ebj,Asakawa:2000tr,Nickel:2006mm,Mueller:2010ah} is an approach that can be used to solve an ill-posed inversion problem, in which the number of data points is much smaller than the number of degrees of freedom available to the function whose reconstruction is sought.  Its basis is Bayes' theorem in probability theory \cite{Jeffreys:1998}, which states the probability of an event ``A'', given that a condition ``B'' is satisfied:
\begin{equation}
P(A|B) = \frac{P(B|A) P(A) }{P(B)}\,,
\end{equation}
where, within the sample space, $P(B|A)$ is the probability that events of type ``A'' satisfy the condition ``B'' (likelihood function); $P(A)$ is the total probability that event ``A'' can occur (prior probability); $P(B)$ is the total probability that condition ``B'' is satisfied (normalisation).

In using the MEM to reconstruct the spectral density, one works with the conditional probability that a spectral function $\rho(\omega)$ corresponds to a correlation function $D(\tau)$:
\begin{equation}
P[\rho | D M ] =\frac{P[D|\rho M]P[\rho|M]}{P[D|M]} \, ,
 \end {equation}
where $M$ represents the body of all definitions and prior knowledge of the spectral function.

According to the central limit theorem, the natural choice for the likelihood functional is
\begin{eqnarray}
P[D|\rho M] & = &\frac{1}{Z_{L}}e^{-L[\rho]}, \\
L[\rho] &=& \sum^{N_{data}}_{i}\frac{(D_{data}(\tau_i)-D_{\rho}(\tau_i))^2}{2\sigma^2_i}\,,
\end{eqnarray}
where $Z_{L}$ is a normalisation factor; $\{D_{data}(\tau_i),i=1,\ldots,N_{data}\}$ are computed from the solution of the gap equation, Eq.\,\eqref{eq:gapeq}, using Eq.\,\eqref{eqDrhoO}; and $\{D_{\rho}(\tau_i),i=1,\ldots,N_{data}\}$ are obtained from Eq.\,\eqref{eq:Drho} using any given model for $\rho(\omega)$.  One typically chooses $\sigma_i = \mathpzc{s}\,D_{data}(\tau_i)$, with $\mathpzc{s} \lesssim 0.01$.

The central feature of the MEM is the prior probability, which is here expressed in terms of the spectral entropy
\begin{equation}
P[\rho|M(\alpha,\mathpzc{m})] = \frac{1}{Z_S}e^{\alpha S[\rho,\mathpzc{m}]} ,
\end{equation}
where $Z_S$ is a normalisation factor, $\alpha$ is a positive-definite scaling factor, and the exponent involves the Shannon-Jaynes entropy \cite{Shannon:1948zz,Jaynes:1957zza,Jaynes:1957zz}
\begin{equation}
S[\rho,\mathpzc{m}] = \int_{-\infty}^{\infty}
\Big[ \rho(\omega)-\mathpzc{m}(\omega)-\rho(\omega) \log\frac{\rho(\omega)}{\mathpzc{m}(\omega)} \Big]\,.
\end{equation}
The quantity $\mathpzc{m}(\omega)$ is the ``default model'' of the spectral function, which is usually chosen to be a uniform distribution so as to avoid assumptions about the structure of the spectral density \cite{Qin:2010pc}; viz.,
\begin{equation}
\label{defaultmodel}
\mathpzc{m}(\omega) = \mathpzc{m}_0 \theta(\Lambda^2 - \omega^2)\,.
\end{equation}
A MEM result for $\rho(\omega)$ is considered reliable if it does not depend on the choices for $\mathpzc{m}_0$, $\Lambda$.

Given a value of $\alpha$, the most probable spectral function, $\rho_\alpha(\omega)$, is obtained by maximising $P[\rho | D M(\alpha,\mathpzc{m}) ]$.  This may be achieved via the singular-value decomposition algorithm in Ref.\,\cite{Bryan:1990ebj}; and dependence on the scale factor $\alpha$ can also be eliminated by following Ref.\,\cite{Bryan:1990ebj} and defining the MEM result for the spectral density as
\begin{eqnarray}
\nonumber
\lefteqn{\bar\rho(\omega) = }\\
&& \int_0^\infty \! \! \! d\alpha \int \!\! \mathpzc{D}\rho(\omega)\, \rho(\omega)\,
P[\rho | D M(\alpha,\mathpzc{m}) ] P[\alpha|D M(\mathpzc{m})]\,,\quad
\label{barrhoomega}
\end{eqnarray}
where the second is a functional integral and $P[\alpha|D M(\alpha,\mathpzc{m})]$ is the conditional probability distribution for $\alpha$.  In readily workable cases, $P[\rho|D M(\alpha,\mathpzc{m})]$ is sharply peaked in the neighbourhood of a single function $\rho_\alpha(\omega)$, in which case Eq.\,\eqref{barrhoomega} yields
\begin{equation}
\label{barrhoomegaA}
\bar\rho(\omega) \approx \int_0^\infty \!\! d\alpha \, \rho_\alpha(\omega)\,P[\alpha|D M(\mathpzc{m})]\,.
\end{equation}

At this point, Bayes' theorem can again be employed to obtain
\begin{equation}
\label{conditionalalpha}
P[\alpha|D M(\mathpzc{m})] \propto \int\mathpzc{D}\rho(\omega)\,
P[\alpha|M(\mathpzc{m})]
P[\rho | D M(\alpha,\mathpzc{m}) ]
\,,
\end{equation}
where we have used the fact that a sensible result is only achieved if it is independent of the default model, Eq.\,\eqref{defaultmodel}.  N.B.\, The conditional probability $P[\alpha|M(\mathpzc{m})]$ is independent of $\rho(\omega)$; and if one considers $\alpha$ and $M$ to be independent, then $P[\alpha|M(\mathpzc{m})]$ is simply a constant.

As noted above, in practically workable instances, $P[\rho|D M(\alpha,\mathpzc{m})]$ is sharply peaked in the neighbourhood of a single function $\rho_\alpha(\omega)$, so the functional integral in Eq.\,\eqref{conditionalalpha} is accurately estimated using Laplace's method, with the result \cite{Bryan:1990ebj}
\begin{eqnarray}
\nonumber
\lefteqn{P[\alpha|D M(\mathpzc{m})]}\\
&\approx& \frac{1}{Z_\Lambda}
\exp\left[\frac{1}{2} \sum_k \ln \left[\frac{\alpha}{\alpha+\lambda_k}\right]\right]
P[\rho_\alpha |D M(\alpha,\mathpzc{m})] \, ,\quad
\label{PalphaDM}
\end{eqnarray}
where $Z_\Lambda$ is a normalisation constant and $\{\lambda_k\}$ is the set of eigenvalues of the real, symmetric matrix
\begin{equation}
\Lambda_{ij}(\rho_\alpha) = \left. \rho_i^{\frac{1}{2}}\frac{\partial^2 L[\rho]}{\partial \rho_i \partial \rho_j} \rho_j^{\frac{1}{2}}\right|_{\rho=\rho_\alpha}
\end{equation}
where the set $\{\rho_i\}$ represents a discretised version of the function $\rho(\omega)$; i.e., the set of values of $\rho(\omega)$ obtained by evaluating the function on a large but finite number of points $\omega \in \mathbb{R}$.

Inserting Eq.\,\eqref{PalphaDM} into Eq.\eqref{barrhoomegaA}, we have our MEM result for the spectral function.

\section{Results and Discussion}
\label{secResults}
\subsection{Critical temperature, $T_c$}
We first solve the gap equation in the chiral limit using each of the six interactions specified in Sec.\,\ref{secTwo} and the interaction parameters in Eq.\,\eqref{eqInteractionParameters}.  Using the solutions and the chiral susceptibility criterion in Ref.\,\cite{Qin:2010nq}, we obtain the critical temperatures for chiral symmetry restoration (in GeV):
\begin{equation}
\begin{array}{lccc}
    & {\rm rainbow} & {\rm BC} & {\rm DB} \\
{\rm MT}  & 0.135 & 0.150 & 0.143\\
{\rm QC}  & 0.142 & 0.160 & 0.149
\end{array}\,.
\end{equation}
Since these values do not differ widely and are obtained using interaction kernels that provide equivalent descriptions of $\pi$ and $\rho$ meson properties, then we judge it reasonable to combine them in order to obtain a DSE prediction for the transition temperature in two-flavour chirally symmetric QCD:
\begin{equation}
T_c = 147 \pm 8\,{\rm MeV},
\label{TcAggressive}
\end{equation}
where the error indicates standard-deviation from the average.  A recent lattice-QCD analysis of the chiral transition temperature in a theory with two massless flavors yields $T_c=154 \pm 9\,$MeV \cite{Bazavov:2011nk}.

Notably, by employing a straightforward generalisation of the inflection point criterion introduced in Refs.\,\cite{Roberts:2007ji,Bashir:2008fk}, one can readily establish that in all cases reflection positivity is violated for $T<T_c$, which signals confinement.  On the other hand, the spectral function is nonnegative for $T>T_c$.  Hence, deconfinement is coincident with chiral symmetry restoration for all interaction kernels in our study.

\begin{figure}[t]
\centerline{\includegraphics[width=0.5\textwidth]{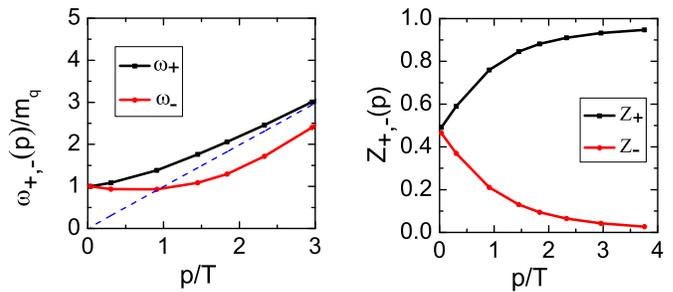}}
\caption{\label{fig:3Tc}
\emph{Left panel} -- quasiparticle dispersion relations at $T=3\,T_c$, where $\omega_{+(-)}$ denotes normal (plasmino) mode.  Diagonal dashed line: free-fermion dispersion relation at this temperature.
\emph{Right panel} -- momentum-dependence of the residues associated with these quasiparticle poles.
Results obtained with rainbow vertex, Eq.\,\protect\eqref{Vrainbow}, and QC interaction, Eq.\,\protect\eqref{DQC}.
}
\end{figure}

\subsection{Far above $T_c$}
\label{FarAboveTc}
With the dressed-quark propagator in hand it is straightforward to determine the MEM spectral function; and so we first checked the approach by computing the spectral function with our six kernels on $T\gtrsim 3\, T_c$.  The results are all qualitatively and semi-quantitatively equivalent. They are also consistent with HTL calculations \cite{Braaten:1990wp}, as apparent in Fig.\,\ref{fig:3Tc}, which shows that our approach yields both a normal and plasmino  mode, with
\begin{equation}
\omega_{\pm}(|\vec{p}|) \stackrel{p\sim 0}{=} m_T \pm 0.33 |\vec{p}|\,.
\end{equation}
The plasmino dispersion law exhibits the expected minimum, in this case at $|\vec{p}|/T\simeq 2/3$; and both $\omega_{\pm}(|\vec{p}|)$ approach free-particle behaviour at $|\vec{p}| \gg T$, with that of the plasmino approaching this limit from below.  The right panel shows that the contribution to the spectral density from the plasmino is strongly damped and contributes little for $p\gtrsim 3\, T$.  These results are in-line with those obtained via simulations of lattice-QCD \cite{Karsch:2009tp}.

It is anticipated from HTL analyses that $T> 0$ propagators exhibit branch cuts whose appearance can be attributed to the opening of scattering channels that are absent at $T=0$ \cite{Weldon:2001vt}; and isolated single quasiparticle poles may broaden to yield a finite lifetime at nonzero temperature \cite{Jeon:1995zm}.
%
In our analysis, however, such branch cuts do not materially contribute to the nonperturbatively-determined spectral density and the spectral peaks are sharp.
This is plausible because a branch point is a lower-order nonanalyticity than a pole; namely, in numerical studies, poles are features with large height, very narrow width and significant spectral strength, whilst branch points are low, broad features with lesser spectral strength.  Thus, compared with poles, branch points may be invisible to a given numerical procedure.  This is similarly true for a small but nonzero width.  Uncovering these features would probably require fine tuning within the MEM, or any other method.
%


Subject to these observations, we have checked whether our numerical results omits significant spectral strength by testing two sum rules for the spectral function:
\begin{eqnarray}
\label{SR1} && Z_{2}^{} \int^{\infty}_{-\infty}\frac{d{\omega^\prime}}{2\pi}\rho_{\pm}(|\vec{p}|,\omega^\prime)=1 ,\\
\label{SR2}
&<\omega>&:=\frac{Z^2_2}{Z^A_2}\int^{\infty}_{-\infty} \frac{d\omega^\prime}{2\pi}\omega^\prime\rho_{\pm}(|\vec{p}|,\omega^\prime)=|\vec{p}|.
\end{eqnarray}

\begin{figure}[t]
\centerline{\includegraphics[width=0.45\textwidth]{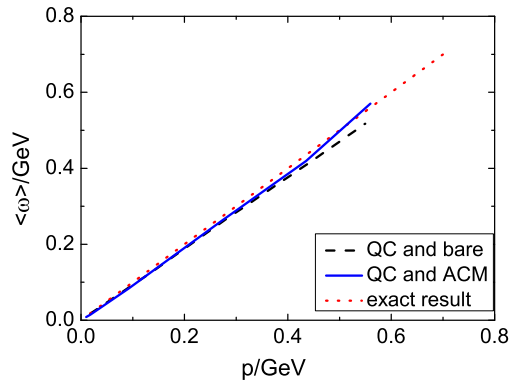}}
\caption{\label{fig:relation}
Numerical check of the momentum sum rule in Eq.\,\protect\eqref{SR2}:
\emph{solid} curve -- DB vertex, Eq.\,\protect\eqref{eq:ACMvertex}, with QC interaction, Eq.\,\protect\eqref{DQC};
\emph{dashed} -- rainbow vertex, Eq.\,\protect\eqref{Vrainbow}, plus QC;
\emph{dotted} -- ideal reference result. The curves would be indistinguishable with a perfect reconstruction of the spectral density.}
\end{figure}

Regarding Eq.\,\eqref{SR1}, we find that $\forall T>T_c$
\begin{equation}
Z_{2}^{} \int^{\infty}_{-\infty}\frac{d\omega^\prime}{2\pi}\rho_{\pm}(|\vec{p}|,\omega^\prime) = Z_{2}^{} \sum_{Q=+,-}Z_{Q}^{}  \, ;
\end{equation}
namely, the sum rule is saturated by the residues of the normal and plasmino modes, a result that Fig.\,\ref{fig:3Tc} illustrates for $T=3\,T_c$.  We discuss the case of $T\gtrsim T_c$ in Sec.\,\ref{secNearTc}.

\begin{figure}[t]
\centerline{\includegraphics[width=0.5\textwidth]{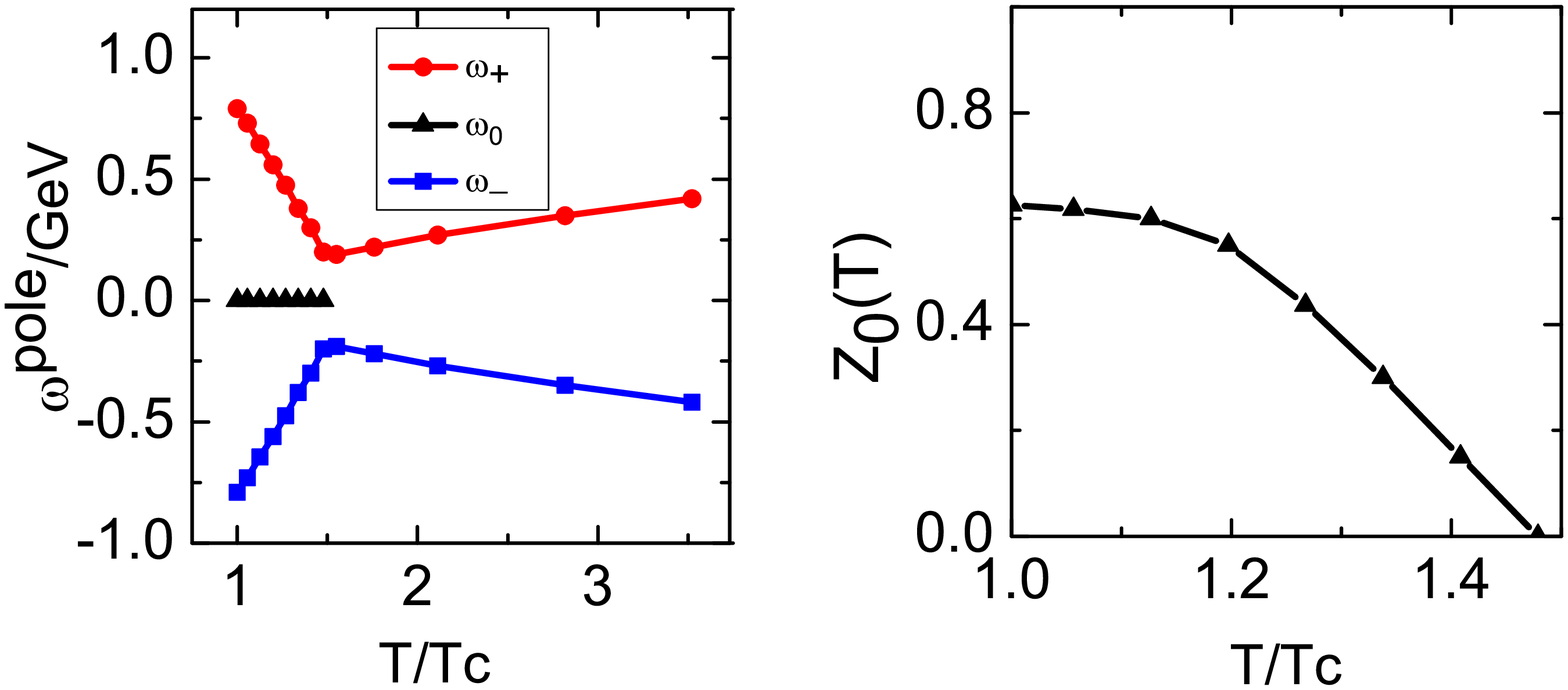}}
\centerline{\includegraphics[width=0.5\textwidth]{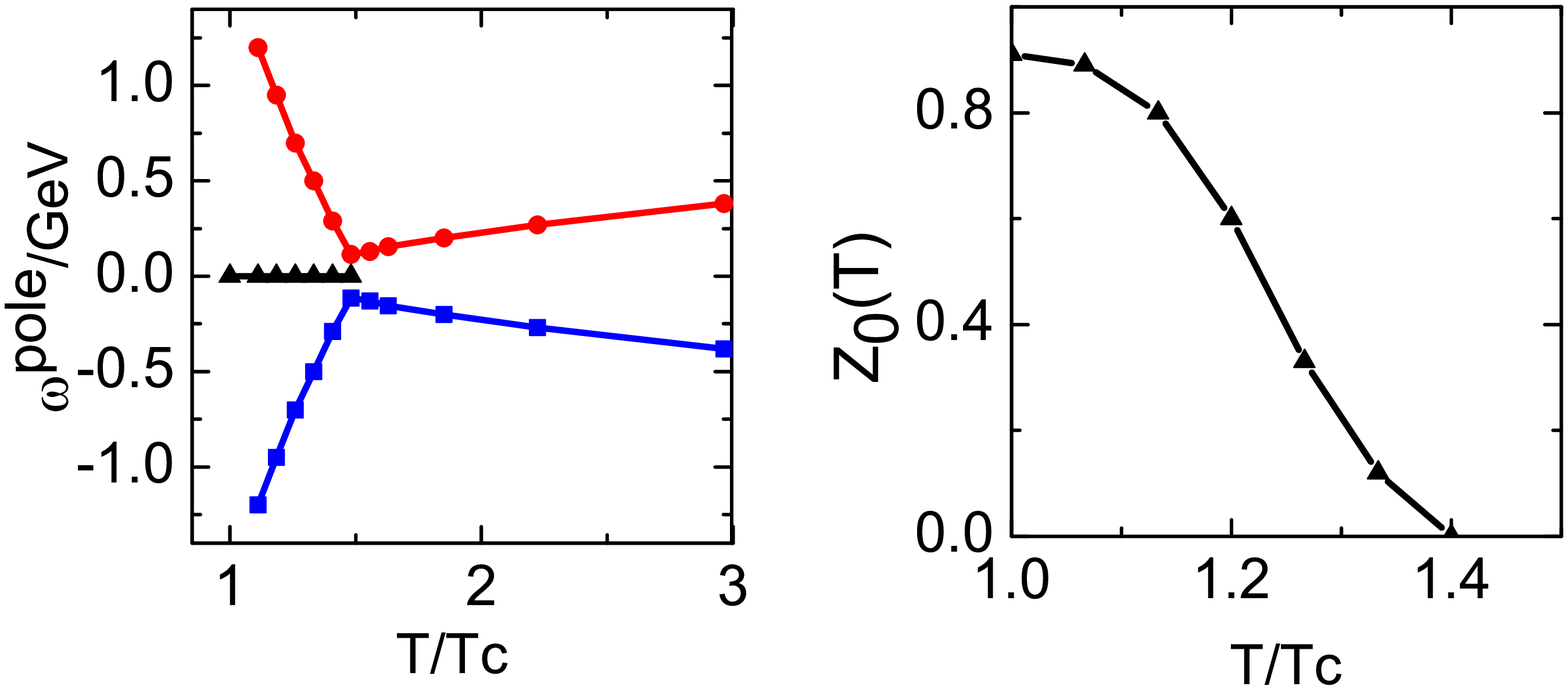}}
\centerline{\includegraphics[width=0.5\textwidth]{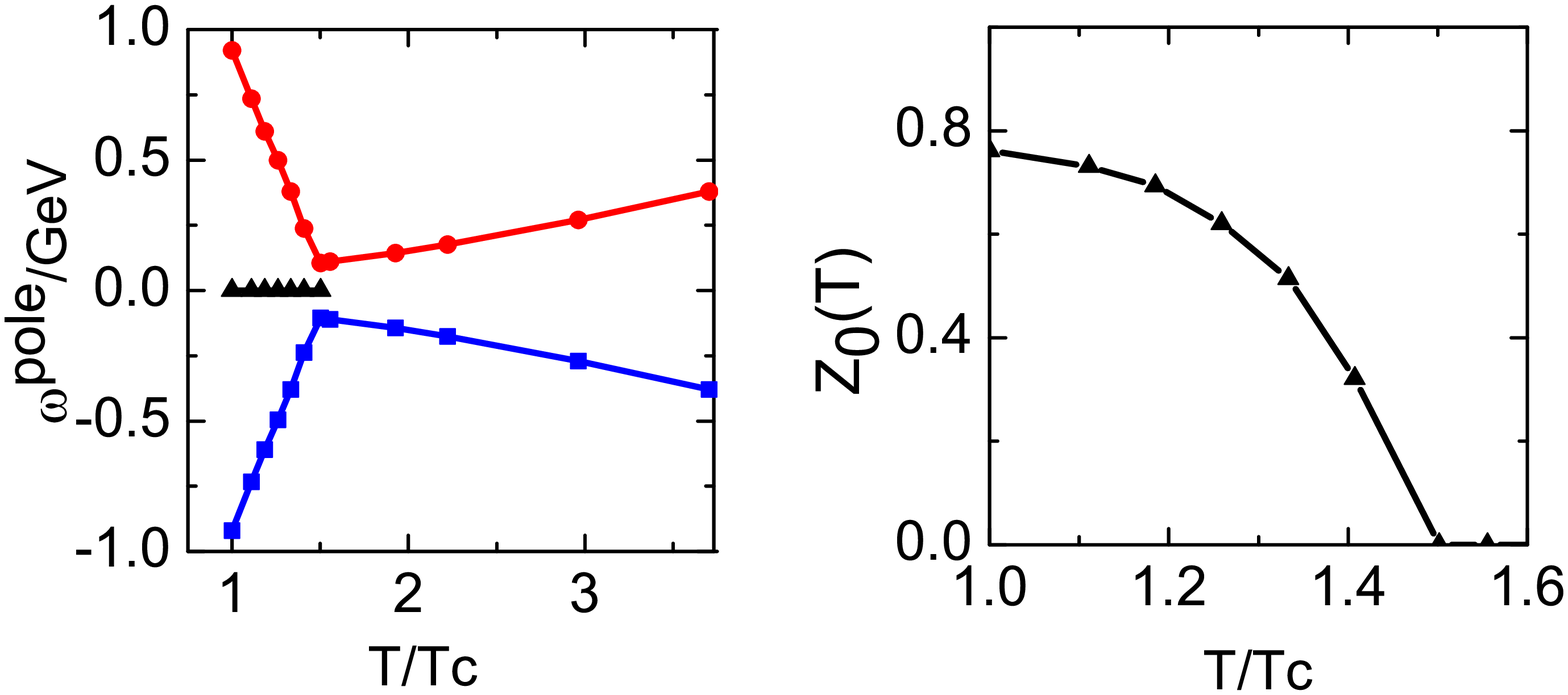}}
\caption{\label{fig:poles}
\emph{Left panels} -- Temperature-dependence of the dressed-quark thermal masses.  Notably, spectral strength is associated with a massless-mode.
\emph{Right panels} -- $T$-dependence of the residue associated with that zero mode.
From top to bottom: QC interaction and rainbow vertex; MT$\,+\,$BC; and QC$\,+\,$DB.}
\end{figure}

The second sum rule, Eq.\,\eqref{SR2}, is readily checked.  We illustrate the result for $T=3\,T_c$ in Fig.\,\ref{fig:relation} using the QC interaction with the rainbow and DB vertices:  the sum rule is satisfied to a high degree of accuracy, limited only by the error on our MEM extraction of the spectral function.  The results from the other kernels are semi-quantitatively equivalent.

\subsection{Neighbourhood of $T_c$}
\label{secNearTc}
In Fig.\,\ref{fig:poles} we depict the $T>T_c$-dependence of the locations of the poles in $\rho(\vec{p}=0,\omega)$; i.e., the thermal masses.  Plainly, as also found for temperatures significantly larger than $T_c$, spectral strength is located at $\omega_+(\vec{p}=0)$ and $\omega_-(\vec{p}=0) = - \omega_+(\vec{p}=0)$, corresponding to the fermion's normal and plasmino modes.  Notable, however, is that on a measurable $T$-domain, spectral strength is also associated with a quasiparticle excitation described by $\omega_0(\vec{p}=0) = 0$.  The appearance of this zero mode is an essentially nonperturbative effect.  It was previously found using the MT interaction in the rainbow truncation \cite{Qin:2010pc}.  This mode is an outgrowth of the evolution in-medium of the gap equation's $T=0$ Wigner-type solution and its appearance here is analogous to the Wigner solution's persistence at nonzero current-quark mass in vacuum \cite{Zong:2004nm,Chang:2006bm,Williams:2006vva,Fischer:2008sp,Wang:2012me,Cui:2013tva}.

We find that with all interaction kernels, the spectral density possesses support associated with this zero mode on $T\in [0,T_s]$, where the value of $T_s$ depends mildly on the interaction, changing by no more than 6\% over the collection we have considered.  As found previously \cite{Qin:2010pc} with the rainbow+$MT$ kernel, Eqs.\,\eqref{Vrainbow} and \eqref{DMT}, all the Wigner-phase spectral strength is located within this mode at $T=0$; it is the dominant contribution for $T\gtrsim T_c$; and, while it is dominant, it is the system's longest wavelength collective mode.
On the other hand, as evident in the right panel of Fig.\,\ref{fig:poles}, the mode's spectral strength diminishes uniformly with increasing $T$ and finally vanishes at
\begin{equation}
T_s/T_c = 1.42 \pm 0.07\,.
\label{TsAggressive}
\end{equation}
Then, for $T>T_s$ the quark's normal and plasmino modes exhibit behavior that is broadly consistent with HTL calculations, as described in Sec.\,\ref{FarAboveTc}.  Most notable, perhaps, is that the thermal masses associated with these modes are a little larger in magnitude when the QC interaction is used.
Given these observations and their lack of material sensitivity to the interaction kernel, we judge that the system should be considered as a sQGP for $T\in [T_c,T_s]$, whereupon it contains a long-range collective mode.

One caveat should be borne in mind when considering this assessment; viz., all interaction kernels we have considered lie within the mean-field class.  The effect of resonant contributions to the gap equation kernel \cite{Holl:1998qs,Fischer:2007ze,Cloet:2008fw,Chang:2009ae} should be checked.  However, so long as such contributions do not materially affect the $T=0$ Wigner-mode solution of the gap equation, and there is no sign that they do, then our conclusions should qualitatively be unaffected.

\begin{figure}[t]
\centerline{\includegraphics[width=0.5\textwidth]{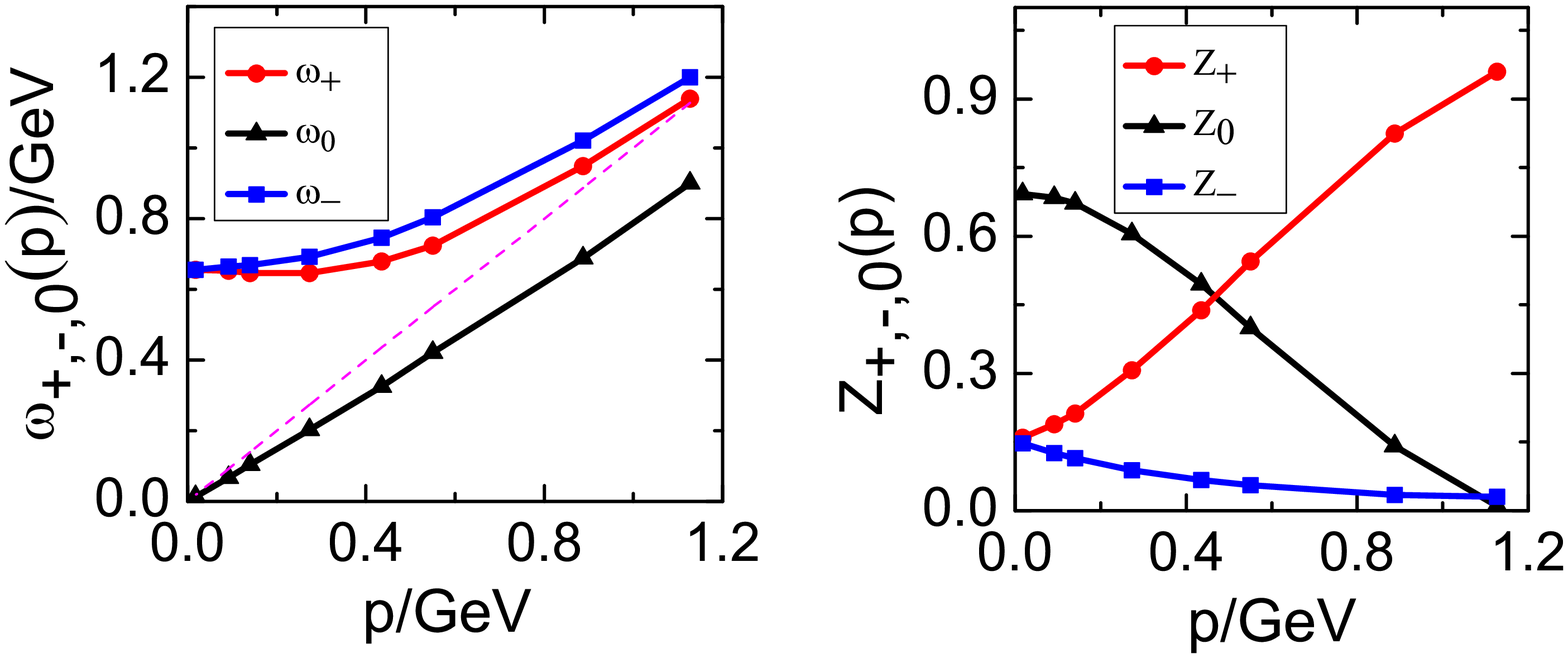}}
\centerline{\includegraphics[width=0.5\textwidth]{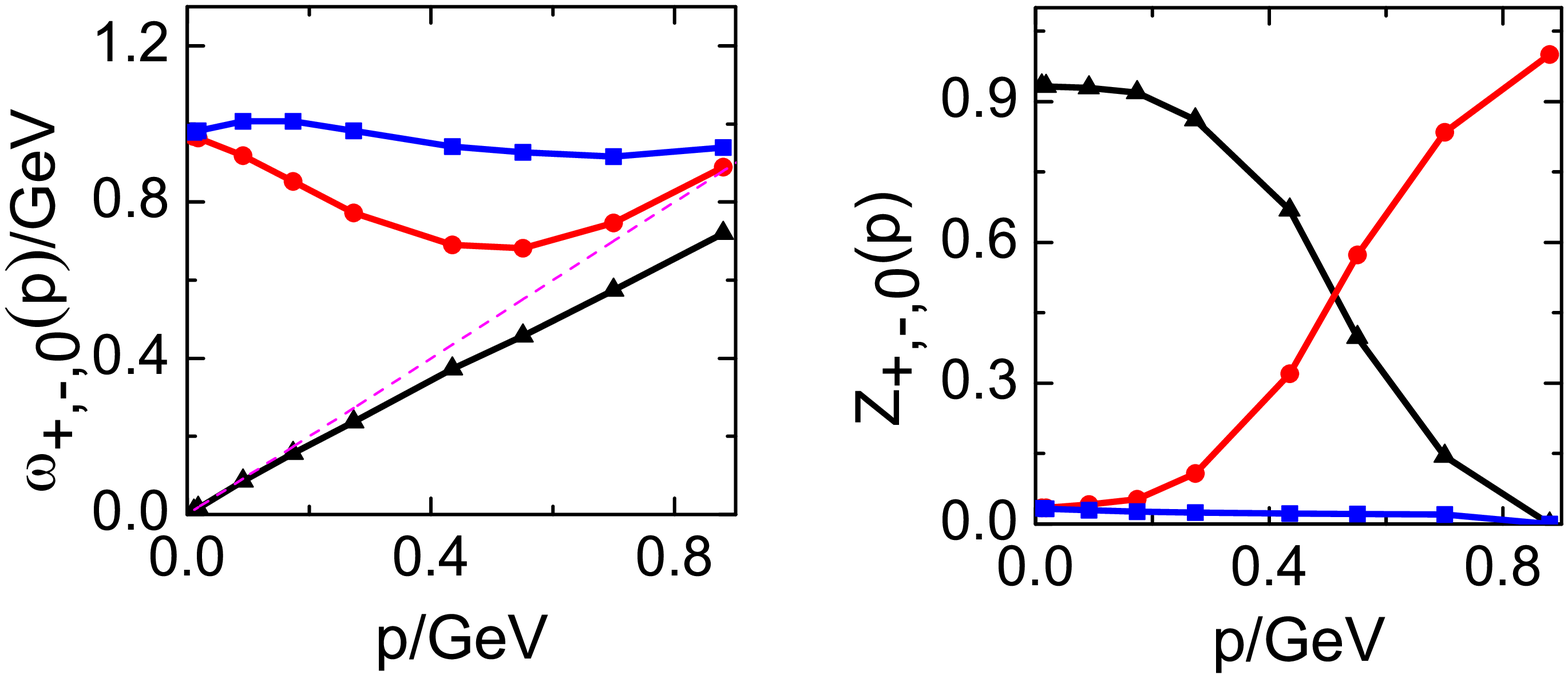}}
\centerline{\includegraphics[width=0.5\textwidth]{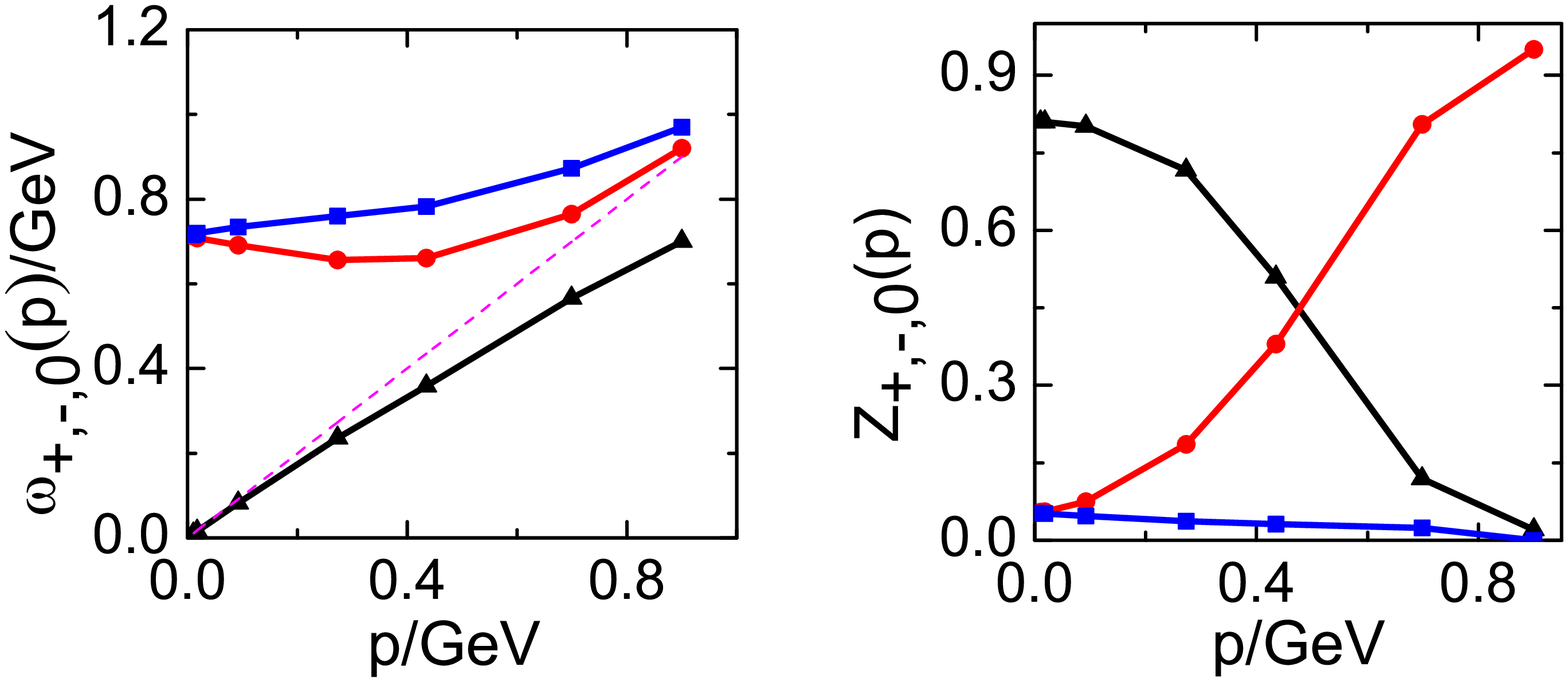}}
%
\caption{\label{fig:dispersion} \emph{Left panels} -- Calculated dispersion relations, $\omega_{\pm,0}$, for all quasiparticles at $T=1.1\,T_c$; and \emph{right panels} -- momentum dependence of the residues of the poles.
As in Fig.\,\ref{fig:poles}, from top to bottom: QC interaction and rainbow vertex; MT$\,+\,$BC; and QC$\,+\,$DB.}
%
%
\end{figure}

In Fig.\,\ref{fig:dispersion} we display the dispersion relations for each quasiparticle mode and the momentum dependence of their residues at $T=1.1\,T_c$.  The results are qualitatively and semi-quantitatively similar in all cases.
The zero mode dominates the spectral density at lower momenta but this role passes to the normal mode at larger momenta.  The momentum at which the switch takes place depends weakly on the gap equation's kernel: it is approximately $0.5\,$GeV for the rainbow and BC vertices, Eqs.\,\eqref{Vrainbow} and \eqref{eq:BCvertex}, and approximately $0.6\,$GeV for the DB vertex, Eq.\,\eqref{eq:ACMvertex}.
The $|\vec{p}|=0$ values of the normal and plasmino mode residues are identical in all cases but the plasmino's residue, and hence its contribution to the spectral density, vanishes quickly with increasing momentum.  Correlated with this, the ``effective mass'' of the plasmino mode; i.e., $\omega_-(|\vec{p}|)$, is always greater than that of the normal mode for $|\vec{p}|>0$.  In the absence of the zero mode; i.e., for $T>T_s$, this pattern is reversed (see, e.g., Fig.\,\ref{fig:3Tc}).

It is notable, and also a useful check on our results, that for each value of $|\vec{p}|$ the sum rule in Eq.\,\eqref{SR1} is saturated, within MEM numerical error, by the addition of all residues.  For example, with the DB vertex, Eq.\,\eqref{eq:ACMvertex}, and QC interaction, Eq.\,\eqref{DQC}, (bottom right panel of Fig.\,\ref{fig:dispersion}):
\begin{equation}
\begin{array}{ccccc}
|\vec{p}|/{\rm GeV}
     & Z_+  & Z_0   & Z_-  & \Sigma_{i=+,0,-} Z_i \\
0.10 & 0.13 & 0.75  & 0.11 & 0.99 \\
0.27 & 0.21 & 0.71  & 0.09 & 1.01 \\
0.57 & 0.45 & 0.50  & 0.06 & 1.01
\end{array}\,.
\end{equation}

We note that Eq.\,(\ref{DTfunction}) is a model and it is natural to enquire after its influence.  None of our results are qualitatively altered by varying $T_{\rm p}$ but, as one would expect, the width of the sQGP domain expands slowly with increasing $T_{\rm p}$; e.g., a 50\% increase in $T_{\rm p}$ typically produces a roughly 30\% increase in $T_s$.



%
\begin{table}[t]
\centering
\caption{Parameter- and model-dependence of the critical temperature for chiral symmetry restoration and deconfinement, $T_c$, and the upper bound of the zero mode's survival domain, $T_s$.  The parameters and kernels are defined in Eqs.\,\protect\eqref{Vrainbow}, \protect\eqref{eq:BCvertex}, \protect\eqref{eq:ACMvertex} and Eqs.\,\protect\eqref{DMT}, \protect\eqref{DQC}.
\label{tab:paradep}}
\begin{tabular}{cccccc}
\hline\hline
Vertex & Interaction & $\, (D\mathpzc{w})^{\frac{1}{3}}/{\textrm{GeV}}$ & $\, \mathpzc{w}/{\textrm{GeV}}$ & $\, T_{c}/{\textrm{MeV}}$ & $\, T_{s}/T_{c}$ \\
\hline
Rainbow & QC  & 0.80 & 0.50 & 142 & 1.55 \\
    & & 0.80 & 0.55 & 132 & 1.59 \\
    & & 0.80 & 0.60 & 123 & 1.67 \\
    & & 0.80 & 0.65 & 118 & 1.70 \\
\hline
Rainbow & MT  & 0.72 & 0.40 & 135 & 1.38 \\
    & & 0.72 & 0.45 & 125 & 1.44 \\
    & & 0.72 & 0.50 & 119 & 1.51 \\
    & & 0.72 & 0.55 & 108 & 1.60 \\
\hline
BC & QC  & 0.60 & 0.50 & 160 & 1.38 \\
    & & 0.62 & 0.55 & 152 & 1.40 \\
    & & 0.64 & 0.60 & 138 & 1.45 \\
    & & 0.66 & 0.65 & 123 & 1.57 \\
\hline
BC & MT  & 0.54 & 0.40 & 150 & 1.40 \\
    & & 0.56 & 0.45 & 135 & 1.46 \\
    & & 0.58 & 0.50 & 123 & 1.60 \\
    & & 0.60 & 0.55 & 110 & 1.75 \\
\hline
DB & QC  & 0.54 & 0.50 & 148 & 1.41\\
    & & 0.57 & 0.55 & 142 & 1.44\\
    & & 0.60 & 0.60 & 132 & 1.52 \\
    & & 0.63 & 0.65 & 121 & 1.69 \\
\hline
DB & MT  & 0.40 & 0.40 & 142 & 1.42 \\
    & & 0.43 & 0.45 & 133 & 1.46 \\
    & & 0.46 & 0.50 & 127 & 1.53 \\
    & & 0.49 & 0.55 & 114 & 1.72 \\
\hline
\hline
\end{tabular}
\end{table}

\subsection{Survey of interaction kernels}
\begin{figure}[t]
\centerline{\includegraphics[width=0.45\textwidth]{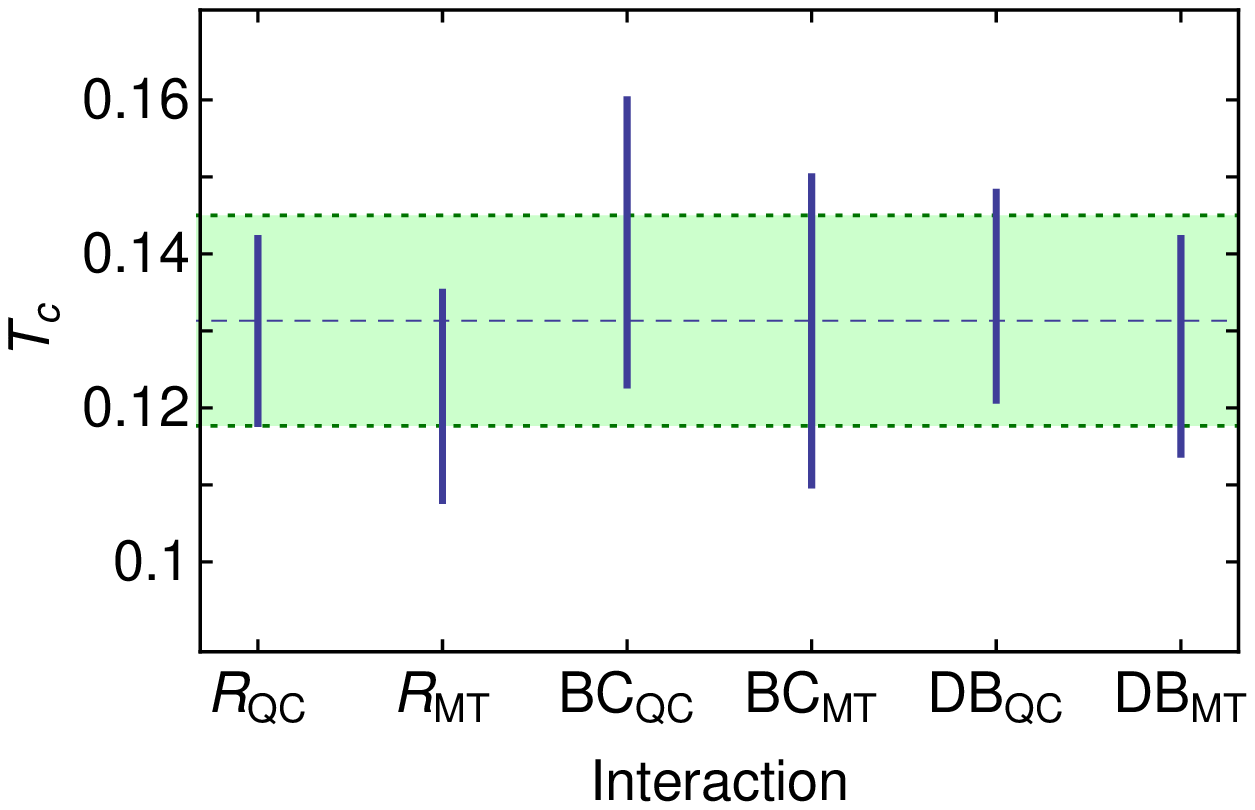}}
\centerline{\includegraphics[width=0.45\textwidth]{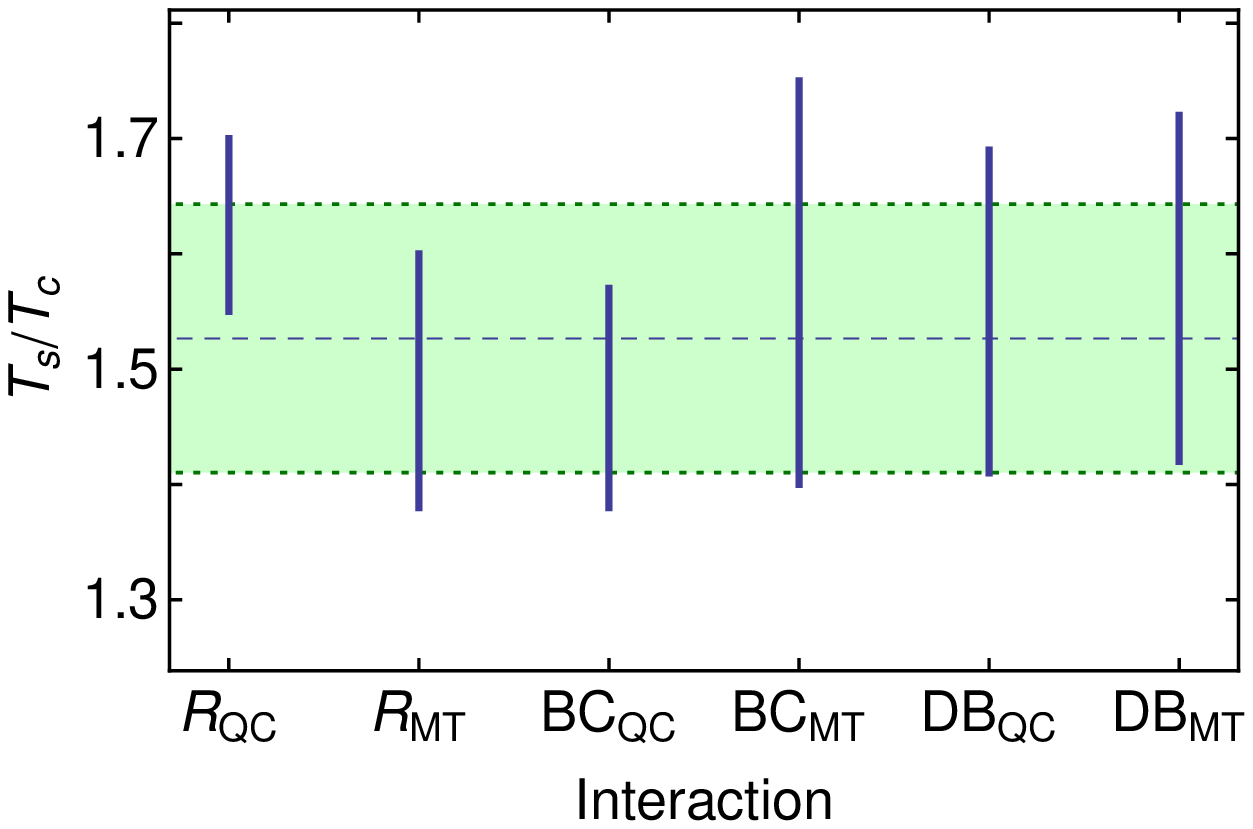}}
\caption{\label{fig:TcTs}
Pictorial representation of results listed in Table~\protect\ref{tab:paradep}.
\emph{Upper panel} -- Interaction and parameter dependence of the critical temperature for chiral symmetry restoration and deconfinement.  (The dashed line and band mark $T_c=0.13\pm 0.014$.)
\emph{Lower panel} -- Analogue for upper bound on the zero mode's survival domain, $T_s$, measured with respect to $T_c$. (The dashed line and band mark $T_s/T_c=1.53\pm 0.12$.)
}
\end{figure}

In order to complete our investigation, we computed the critical temperature for chiral symmetry restoration and the zero mode's survival domain in each model on that domain of interaction strengths which preserves a uniform value for the chiral-limit value of the in-pion condensate; viz., $\langle \bar q q \rangle_{1\,{\rm GeV}}^\pi \approx (-0.24\,{\rm GeV})^3$.  The results are listed in Table~\ref{tab:paradep} and depicted in Fig.\,\ref{fig:TcTs}.

The trends are clear and uniform across all six models.  Namely, the relative width of the survival domain increases as the absolute value of the critical temperature diminishes; and, as observed elsewhere \cite{Qin:2010nq}, $T_c$ decreases as the confinement length-scale $r_\mathpzc{w} = 1/\mathpzc{w}$ decreases.

The panels in Fig.\,\ref{fig:TcTs} show that the variation domains for each interaction overlap considerably and hence it is sensible to list their averages as conservative measures of the critical temperature and zero mode survival domain:
\begin{equation}
\label{TcConservative}
T_c = 0.131\pm 0.014\,{\rm GeV},\quad
T_s/T_c = 1.53 \pm 0.12\,.
\end{equation}
These values are consistent with those in Eqs.\,\eqref{TcAggressive}, \eqref{TsAggressive}.  The latter describe estimates based on parameter values at the midpoint of the domain within which zero temperature observables show almost no sensitivity to parameter variation (see the introduction to Sec.\,\ref{secQualitative}).  (N.B.\ The uncertainty on $T_s$ does not include a response to variations of $T_{\rm p}$ in Eq.\,\eqref{DTfunction}.  Should good reason be offered for $T_{\rm p} > T_c$, then, as remarked already, this will modestly increase the central value of $T_s/T_c$.)

\subsection{Wider view}
Additional context for our results is provided by observing that the appearance of a third and long-wavelength mode in the dressed-fermion spectral density on a material temperature domain above $T_c$ has also been seen in straightforward one-loop computations of the fermion self-energy, irrespective of the nature of the boson which dresses the fermion
\cite{Kitazawa:2005mp,Kitazawa:2007ep,Harada:2008vk}.  Where a comparison is possible, the dependence of our spectral density on $(\omega,|\vec{p}|,T)$ is similar to that seen in one-loop analyses of model gap equations.  In analogy with a similar effect in high-temperature superconductivity \cite{janko:1997}, that behaviour has been attributed to Landau damping, an interference phenomenon known from plasma physics.  Indeed, Landau damping is typical of in-medium self-energy corrections when the thermal energy of the fermion, $\pi T$, is commensurate with the mass-scale which characterises the dispersion law of the dressing boson; viz., $M_g \approx 0.5\,{\rm GeV}$ in QCD (see, e.g., Sec.\,2.3 in Ref.\,\cite{Cloet:2013jya}).

Notably, our analysis shows that a coupling to meson-like correlations in the gap equation is not a precondition for appearance of the zero mode because such correlations are absent in mean-field truncations \cite{Holl:1998qs}.  On the other hand, our gap equation's kernel is characterised by an interaction that features an infrared mass-scale $M_g \approx \pi T_c$ and supports dynamical chiral symmetry breaking in-vacuum.  We anticipate that the zero mode will markedly affect colour-singlet vacuum polarisations on $T\in [T_c,T_s]$.  This could be explicated using the methods of Refs.\,\cite{Chang:2008ec,Chang:2009at}.

Our results, in conjunction with these remarks describing their compatibility with outcomes of other studies whose foundations are distinct, suggest strongly that the existence of a three-peak spectral density, with a zero mode, on a domain that extends to approximately $1.5\,T_c$, is a feature of any realistic kernel for QCD's gap equation.


\section{Concluding Remarks}
\label{secEnd}
In order to expose the active fermionic quasiparticle degrees-of-freedom in the neighbourhood of the critical temperature for chiral symmetry restoration in massless two-flavour QCD, $T_c$, we analysed the phase transition obtained with three distinct gluon-quark vertices and two different assumptions about the long-range part of the quark-quark interaction.  With each of the six permutations that these inputs allow, we solved the gap equation, located the transition temperature $T_c$, and used the maximum entropy method (MEM) to extract the dressed-quark spectral function at $T>T_c$.

Each of the kernels is characterised by a single parameter, which may be interpreted as defining a length-scale for confinement, $r_\mathpzc{w}$; and numerous in-vacuum observables are independent of this parameter for variations of $\pm\, 20$\% about its optimal value.  Based on these observations, we obtained a best estimate for the critical temperature; viz., $T_c=147\pm 8\,$MeV, described in connection with Eq.\,\eqref{TcAggressive}.  In this study, as with those DSE analyses preceding it, deconfinement at nonzero temperature is coincident with chiral symmetry restoration.  If we allow for variations in $r_\mathpzc{w}$ within the domain of in-vacuum stability, then we obtain the more conservative estimate: $T_c=131\pm 14\,$MeV, discussed in connection with Eq.\,\eqref{TcConservative}.

We demonstrated that the MEM is a reliable tool for reconstructing the quark spectral density and therewith obtained a result for that density which is consistent with those produced using a hard thermal loop expansion at temperatures markedly above $T_c$, exhibiting both a normal and plasmino mode (see Sec.\,\ref{FarAboveTc}).

On the other hand, with each of the six kernels we considered, the spectral function contains a significant additional feature on a domain $T\in [T_c,T_s]$, with $T_s \simeq 1.5 T_c$.  Therein, as discussed in Sec.\,\ref{secNearTc}, the spectral function displays a third peak, associated with a zero mode; i.e., a long-wavelength quasiparticle mode described by a dispersion law $\omega_0(|\vec{p}|)$, with $\omega_0(0)=0$.  This mode is essentially nonperturbative in origin and dominates the spectral function at $T=T_c$.  Our best estimate for the upper bound of its survival domain is $T_s/T_c = 1.42 \pm 0.07$, as described in connection with Eq.\,\eqref{TsAggressive}.  A more conservative estimate, accommodating variations in $r_\mathpzc{w}$, is $T_s/T_c = 1.53 \pm 0.12$, explained in connection with Eq.\,\eqref{TcConservative}.

Notwithstanding the fact that all interaction kernels we considered lie within the mean-field class, we presented arguments and examples that suggest it should survive the inclusion of resonant contributions to the gap equation kernel.  We therefore judge that the existence of this mode is a signal for the formation of a strongly-coupled QGP and, moreover, that this strongly-interacting state of matter is probably a distinctive feature of the QCD phase transition.


\section*{Acknowledgments}
SXQ acknowledges support from the Alexander von Humboldt Foundation via a Research Fellowship for Postdoctoral Researchers, CDR acknowledges support from an \emph{International Fellow Award} from the Helmholtz Association, and this work was otherwise supported by:
the National Natural Science Foundation of China under Contract Nos.\ 10935001, 11075052 and 11175004;
the National Key Basic Research Program of China under Contract No.\ G2013CB834400;
the U.\,S.\ Department of Energy, Office of Nuclear Physics, Contract no.~DE-AC02-06CH11357;
and For\-schungs\-zentrum J\"ulich GmbH.



\end{document}